

%
%
%
\def\unredoffs{} \def\redoffs{\voffset=-.31truein\hoffset=-.59truein}
\def\speclscape{\special{ps: landscape}}
%
%
%
%
\newbox\leftpage \newdimen\fullhsize \newdimen\hstitle \newdimen\hsbody
\tolerance=1000\hfuzz=2pt
\catcode`\@=11 
\def\bigans{b }
\def\answ{b }

%
\ifx\answ\bigans\message{(This will come out unreduced.}
\magnification=1200\unredoffs\baselineskip=16pt plus 2pt minus 1pt
\hsbody=\hsize \hstitle=\hsize 
\else\message{(This will be reduced.} \let\l@r=L
\magnification=1000\baselineskip=16pt plus 2pt minus 1pt \vsize=7truein
\redoffs \hstitle=8truein\hsbody=4.75truein\fullhsize=10truein\hsize=\hsbody
\output={\ifnum\pageno=0 
  \shipout\vbox{\speclscape{\hsize\fullhsize\makeheadline}
    \hbox to \fullhsize{\hfill\pagebody\hfill}}\advancepageno
  \else
  \almostshipout{\leftline{\vbox{\pagebody\makefootline}}}\advancepageno
  \fi}
\def\almostshipout#1{\if L\l@r \count1=1 \message{[\the\count0.\the\count1]}
      \global\setbox\leftpage=#1 \global\let\l@r=R
 \else \count1=2
  \shipout\vbox{\speclscape{\hsize\fullhsize\makeheadline}
      \hbox to\fullhsize{\box\leftpage\hfil#1}}  \global\let\l@r=L\fi}
\fi
%
\newcount\yearltd\yearltd=\year\advance\yearltd by -1900

\def\Title#1#2{\nopagenumbers\abstractfont\hsize=\hstitle\rightline{#1}%
\vskip 1in\centerline{\titlefont #2}\abstractfont\vskip .5in\pageno=0}
\def\Date#1{\vfill\leftline{#1}\tenpoint\supereject\global\hsize=\hsbody%
\footline={\hss\tenrm\folio\hss}}
%

\def\draftmode{\message{ DRAFTMODE }\def\draftdate{{\rm preliminary draft:
\number\month/\number\day/\number\yearltd\ \ \hourmin}}%
\headline={\hfil\draftdate}\writelabels\baselineskip=20pt plus 2pt minus 2pt
 {\count255=\time\divide\count255 by 60 \xdef\hourmin{\number\count255}
  \multiply\count255 by-60\advance\count255 by\time
  \xdef\hourmin{\hourmin:\ifnum\count255<10 0\fi\the\count255}}}
\def\nolabels{\def\wrlabeL##1{}\def\eqlabeL##1{}\def\reflabeL##1{}}
\def\writelabels{\def\wrlabeL##1{\leavevmode\vadjust{\rlap{\smash%
{\line{{\escapechar=` \hfill\rlap{\sevenrm\hskip.03in\string##1}}}}}}}%
\def\eqlabeL##1{{\escapechar-1\rlap{\sevenrm\hskip.05in\string##1}}}%
\def\reflabeL##1{\noexpand\llap{\noexpand\sevenrm\string\string\string##1}}}
\nolabels
%
\global\newcount\secno \global\secno=0
\global\newcount\meqno \global\meqno=1
\def\newsec#1{\global\advance\secno by1\message{(\the\secno. #1)}
\global\subsecno=0\eqnres@t\noindent{\bf\the\secno. #1}
\writetoca{{\secsym} {#1}}\par\nobreak\medskip\nobreak}
\def\eqnres@t{\xdef\secsym{\the\secno.}\global\meqno=1\bigbreak\bigskip}
\def\sequentialequations{\def\eqnres@t{\bigbreak}}\xdef\secsym{}
\global\newcount\subsecno \global\subsecno=0
\def\subsec#1{\global\advance\subsecno by1\message{(\secsym\the\subsecno. #1)}
\ifnum\lastpenalty>9000\else\bigbreak\fi
\noindent{\it\secsym\the\subsecno. #1}\writetoca{\string\quad
{\secsym\the\subsecno.} {#1}}\par\nobreak\medskip\nobreak}
\def\appendix#1#2{\global\meqno=1\global\subsecno=0\xdef\secsym{\hbox{#1.}}
\bigbreak\bigskip\noindent{\bf Appendix #1. #2}\message{(#1. #2)}
\writetoca{Appendix {#1.} {#2}}\par\nobreak\medskip\nobreak}
%
%
\def\eqnn#1{\xdef #1{(\secsym\the\meqno)}\writedef{#1\leftbracket#1}%
\global\advance\meqno by1\wrlabeL#1}
\def\eqna#1{\xdef #1##1{\hbox{$(\secsym\the\meqno##1)$}}
\writedef{#1\numbersign1\leftbracket#1{\numbersign1}}%
\global\advance\meqno by1\wrlabeL{#1$\{\}$}}
\def\eqn#1#2{\xdef #1{(\secsym\the\meqno)}\writedef{#1\leftbracket#1}%
\global\advance\meqno by1$$#2\eqno#1\eqlabeL#1$$}
%
\newskip\footskip\footskip14pt plus 1pt minus 1pt 
\def\footnotefont{\ninepoint}\def\f@t#1{\footnotefont #1\@foot}
\def\f@@t{\baselineskip\footskip\bgroup\footnotefont\aftergroup\@foot\let\next}
\setbox\strutbox=\hbox{\vrule height9.5pt depth4.5pt width0pt}
\global\newcount\ftno \global\ftno=0
\def\foot{\global\advance\ftno by1\footnote{$^{\the\ftno}$}}
%
\newwrite\ftfile
\def\footend{\def\foot{\global\advance\ftno by1\chardef\wfile=\ftfile
$^{\the\ftno}$\ifnum\ftno=1\immediate\openout\ftfile=foots.tmp\fi%
\immediate\write\ftfile{\noexpand\smallskip%
\noexpand\item{f\the\ftno:\ }\pctsign}\findarg}%
\def\footatend{\vfill\eject\immediate\closeout\ftfile{\parindent=20pt
\centerline{\bf Footnotes}\nobreak\bigskip\input foots.tmp }}}
\def\footatend{}
%
%
\global\newcount\refno \global\refno=1
\newwrite\rfile
\def\ref{[\the\refno]\nref}
\def\nref#1{\xdef#1{[\the\refno]}\writedef{#1\leftbracket#1}%
\ifnum\refno=1\immediate\openout\rfile=refs.tmp\fi
\global\advance\refno by1\chardef\wfile=\rfile\immediate
\write\rfile{\noexpand\item{#1\ }\reflabeL{#1\hskip.31in}\pctsign}\findarg}
\def\findarg#1#{\begingroup\obeylines\newlinechar=`\^^M\pass@rg}
{\obeylines\gdef\pass@rg#1{\writ@line\relax #1^^M\hbox{}^^M}%
\gdef\writ@line#1^^M{\expandafter\toks0\expandafter{\striprel@x #1}%
\edef\next{\the\toks0}\ifx\next\em@rk\let\next=\endgroup\else\ifx\next\empty%
\else\immediate\write\wfile{\the\toks0}\fi\let\next=\writ@line\fi\next\relax}}
\def\striprel@x#1{} \def\em@rk{\hbox{}}
\def\lref{\begingroup\obeylines\lr@f}
\def\lr@f#1#2{\gdef#1{\ref#1{#2}}\endgroup\unskip}

\def\addref#1{\immediate\write\rfile{\noexpand\item{}#1}} 
\def\startrefs#1{\immediate\openout\rfile=refs.tmp\refno=#1}
\def\xref{\expandafter\xr@f}\def\xr@f[#1]{#1}
\def\refs#1{\count255=1[\r@fs #1{\hbox{}}]}
\def\r@fs#1{\ifx\und@fined#1\message{reflabel \string#1 is undefined.}%
\nref#1{need to supply reference \string#1.}\fi%
\vphantom{\hphantom{#1}}\edef\next{#1}\ifx\next\em@rk\def\next{}%
\else\ifx\next#1\ifodd\count255\relax\xref#1\count255=0\fi%
\else#1\count255=1\fi\let\next=\r@fs\fi\next}
%

%
\newwrite\ffile\global\newcount\figno \global\figno=1
\def\fig{fig.~\the\figno\nfig}
\def\nfig#1{\xdef#1{fig.~\the\figno}%
\writedef{#1\leftbracket fig.\noexpand~\the\figno}%
\ifnum\figno=1\immediate\openout\ffile=figs.tmp\fi\chardef\wfile=\ffile%
\immediate\write\ffile{\noexpand\medskip\noexpand\item{Fig.\ \the\figno. }
\reflabeL{#1\hskip.55in}\pctsign}\global\advance\figno by1\findarg}
\def\xfig{\expandafter\xf@g}\def\xf@g fig.\penalty\@M\ {}
\def\figs#1{figs.~\f@gs #1{\hbox{}}}
\def\f@gs#1{\edef\next{#1}\ifx\next\em@rk\def\next{}\else
\ifx\next#1\xfig #1\else#1\fi\let\next=\f@gs\fi\next}
\newwrite\lfile
{\escapechar-1\xdef\pctsign{\string\%}\xdef\leftbracket{\string\{}
\xdef\rightbracket{\string\}}\xdef\numbersign{\string\#}}

\def\writestop{\def\writestoppt{\immediate\write\lfile{\string\pageno%
\the\pageno\string\startrefs\leftbracket\the\refno\rightbracket%
\string\def\string\secsym\leftbracket\secsym\rightbracket%
\string\secno\the\secno\string\meqno\the\meqno}\immediate\closeout\lfile}}
\def\writestoppt{}\def\writedef#1{}
\def\seclab#1{\xdef #1{\the\secno}\writedef{#1\leftbracket#1}\wrlabeL{#1=#1}}
\def\subseclab#1{\xdef #1{\secsym\the\subsecno}%
\writedef{#1\leftbracket#1}\wrlabeL{#1=#1}}
\newwrite\tfile \def\writetoca#1{}
\def\leaderfill{\leaders\hbox to 1em{\hss.\hss}\hfill}
\def\writetoc{\immediate\openout\tfile=toc.tmp
   \def\writetoca##1{{\edef\next{\write\tfile{\noindent ##1
   \string\leaderfill {\noexpand\number\pageno} \par}}\next}}}

%
\catcode`\@=12 
%
\edef\tfontsize{\ifx\answ\bigans scaled\magstep3\else scaled\magstep4\fi}
\font\titlerm=cmr10 \tfontsize \font\titlerms=cmr7 \tfontsize
\font\titlermss=cmr5 \tfontsize \font\titlei=cmmi10 \tfontsize
\font\titleis=cmmi7 \tfontsize \font\titleiss=cmmi5 \tfontsize
\font\titlesy=cmsy10 \tfontsize \font\titlesys=cmsy7 \tfontsize
\font\titlesyss=cmsy5 \tfontsize \font\titleit=cmti10 \tfontsize
\skewchar\titlei='177 \skewchar\titleis='177 \skewchar\titleiss='177
\skewchar\titlesy='60 \skewchar\titlesys='60 \skewchar\titlesyss='60
\def\titlefont{\def\rm{\fam0\titlerm}
\textfont0=\titlerm \scriptfont0=\titlerms \scriptscriptfont0=\titlermss
\textfont1=\titlei \scriptfont1=\titleis \scriptscriptfont1=\titleiss
\textfont2=\titlesy \scriptfont2=\titlesys \scriptscriptfont2=\titlesyss
\textfont\itfam=\titleit \def\it{\fam\itfam\titleit}\rm}
 \ifx\answ\bigans\else scaled\magstep1\fi
\ifx\answ\bigans\def\abstractfont{\tenpoint}\else
\font\abssl=cmsl10 scaled \magstep1
\font\absrm=cmr10 scaled\magstep1 \font\absrms=cmr7 scaled\magstep1
\font\absrmss=cmr5 scaled\magstep1 \font\absi=cmmi10 scaled\magstep1
\font\absis=cmmi7 scaled\magstep1 \font\absiss=cmmi5 scaled\magstep1
\font\abssy=cmsy10 scaled\magstep1 \font\abssys=cmsy7 scaled\magstep1
\font\abssyss=cmsy5 scaled\magstep1 \font\absbf=cmbx10 scaled\magstep1
\skewchar\absi='177 \skewchar\absis='177 \skewchar\absiss='177
\skewchar\abssy='60 \skewchar\abssys='60 \skewchar\abssyss='60
\def\abstractfont{\def\rm{\fam0\absrm}
\textfont0=\absrm \scriptfont0=\absrms \scriptscriptfont0=\absrmss
\textfont1=\absi \scriptfont1=\absis \scriptscriptfont1=\absiss
\textfont2=\abssy \scriptfont2=\abssys \scriptscriptfont2=\abssyss
\textfont\itfam=\bigit \def\it{\fam\itfam\bigit}\def\footnotefont{\tenpoint}%
\textfont\slfam=\abssl \def\sl{\fam\slfam\abssl}%
\textfont\bffam=\absbf \def\bf{\fam\bffam\absbf}\rm}\fi
\def\tenpoint{\def\rm{\fam0\tenrm}
\textfont0=\tenrm \scriptfont0=\sevenrm \scriptscriptfont0=\fiverm
\textfont1=\teni  \scriptfont1=\seveni  \scriptscriptfont1=\fivei
\textfont2=\tensy \scriptfont2=\sevensy \scriptscriptfont2=\fivesy
\textfont\itfam=\tenit \def\it{\fam\itfam\tenit}\def\footnotefont{\ninepoint}%
\textfont\bffam=\tenbf \def\bf{\fam\bffam\tenbf}\def\sl{\fam\slfam\tensl}\rm}
\font\ninerm=cmr9 \font\sixrm=cmr6 \font\ninei=cmmi9 \font\sixi=cmmi6
\font\ninesy=cmsy9 \font\sixsy=cmsy6 \font\ninebf=cmbx9
\font\nineit=cmti9 \font\ninesl=cmsl9 \skewchar\ninei='177
\skewchar\sixi='177 \skewchar\ninesy='60 \skewchar\sixsy='60
\def\ninepoint{\def\rm{\fam0\ninerm}
\textfont0=\ninerm \scriptfont0=\sixrm \scriptscriptfont0=\fiverm
\textfont1=\ninei \scriptfont1=\sixi \scriptscriptfont1=\fivei
\textfont2=\ninesy \scriptfont2=\sixsy \scriptscriptfont2=\fivesy
\textfont\itfam=\ninei \def\it{\fam\itfam\nineit}\def\sl{\fam\slfam\ninesl}%
\textfont\bffam=\ninebf \def\bf{\fam\bffam\ninebf}\rm}
%
%

\hyphenation{anom-aly anom-alies coun-ter-term coun-ter-terms}
\def\inv{^{\raise.15ex\hbox{${\scriptscriptstyle -}$}\kern-.05em 1}}

\def\Dsl{\,\raise.15ex\hbox{/}\mkern-13.5mu D} 
\def\dsl{\raise.15ex\hbox{/}\kern-.57em\partial}

 \def\Tr{{\rm Tr}}
\font\bigit=cmti10 scaled \magstep1
\def\lspace{\ifx\answ\bigans{}\else\qquad\fi}
\def\lbspace{\ifx\answ\bigans{}\else\hskip-.2in\fi} 
\def\boxeqn#1{\vcenter{\vbox{\hrule\hbox{\vrule\kern3pt\vbox{\kern3pt
	\hbox{${\displaystyle #1}$}\kern3pt}\kern3pt\vrule}\hrule}}}
\def\mbox#1#2{\vcenter{\hrule \hbox{\vrule height#2in
		\kern#1in \vrule} \hrule}}  
%
 \def\CC{{\cal C}}

\def\darr#1{\raise1.5ex\hbox{$\leftrightarrow$}\mkern-16.5mu #1}

\def\roughly#1{\raise.3ex\hbox{$#1$\kern-.75em\lower1ex\hbox{$\sim$}}}

\let\includefigures=\iftrue
\let\useblackboard=\iftrue
\newfam\black

\includefigures
\message{If you do not have epsf.tex (to include figures),}
\message{change the option at the top of the tex file.}
\input epsf
\def\figin{\epsfcheck\figin}\def\figins{\epsfcheck\figins}
\def\epsfcheck{\ifx\epsfbox\UnDeFiNeD
\message{(NO epsf.tex, FIGURES WILL BE IGNORED)}
\gdef\figin##1{\vskip2in}\gdef\figins##1{\hskip.5in}
\else\message{(FIGURES WILL BE INCLUDED)}%
\gdef\figin##1{##1}\gdef\figins##1{##1}\fi}
\def\DefWarn#1{}
\def\figinsert{\goodbreak\midinsert}
\def\ifig#1#2#3{\DefWarn#1\xdef#1{fig.~\the\figno}
\writedef{#1\leftbracket fig.\noexpand~\the\figno}%
\figinsert\figin{\centerline{#3}}\medskip\centerline{\vbox{
\baselineskip12pt\advance\hsize by -1truein
\noindent\footnotefont{\bf Fig.~\the\figno:} #2}}
\endinsert\global\advance\figno by1}
\else
\def\ifig#1#2#3{\xdef#1{fig.~\the\figno}
\writedef{#1\leftbracket fig.\noexpand~\the\figno}%
\global\advance\figno by1} \fi

\def\id{{1 \kern-.28em {\rm l}}}

\def\K3{{\bf K3}}
\def\journal#1&#2(#3){\unskip, \sl #1\ \bf #2 \rm(19#3) }
\def\andjournal#1&#2(#3){\sl #1~\bf #2 \rm (19#3) }

\def\tilde{\widetilde}

\def\frac#1#2{{#1\over#2}}

\def\inbar{\,\vrule height1.5ex width.4pt depth0pt}
\def\IC{\relax\hbox{$\inbar\kern-.3em{\rm C}$}}
\def\IR{\relax{\rm I\kern-.18em R}}
\def\IP{\relax{\rm I\kern-.18em P}}

%
%

%
\catcode`\@=11
\def\slash#1{\mathord{\mathpalette\c@ncel{#1}}}
\overfullrule=0pt
\def\AA{{\cal A}}
\def\BB{{\cal B}}
\def\CC{{\cal C}}

\def\LL{{\cal L}}

\def\OO{{\cal O}}

\def\underrel#1\over#2{\mathrel{\mathop{\kern\z@#1}\limits_{#2}}}

\catcode`\@=12


%

\def\det{{\rm det}}

\def\Tr{{\rm Tr}}

\def\det{{\rm det}}


\def\LL{{\cal L}}
\def\tb{{\tilde b}}


\lref\BastianelliFTa{
  F.~Bastianelli, S.~Frolov and A.~A.~Tseytlin,
  ``Three-point correlators of stress tensors in maximally-supersymmetric
  conformal theories in d = 3 and d = 6,''
  Nucl.\ Phys.\  B {\bf 578}, 139 (2000)
  [arXiv:hep-th/9911135].
}

\lref\BastianelliFTb{
  F.~Bastianelli, S.~Frolov and A.~A.~Tseytlin,
  ``Conformal anomaly of (2,0) tensor multiplet in six dimensions and  AdS/CFT
  correspondence,''
  JHEP {\bf 0002}, 013 (2000)
  [arXiv:hep-th/0001041].
}


\lref\HofmanMaldacena{
  D.~M.~Hofman and J.~Maldacena,
  ``Conformal collider physics: Energy and charge correlations,''
  JHEP {\bf 0805}, 012 (2008)
  [arXiv:0803.1467 [hep-th]].
}

\lref\deBoerPN{
  J.~de Boer, M.~Kulaxizi and A.~Parnachev,
  ``$AdS_7/CFT_6$, Gauss-Bonnet Gravity, and Viscosity Bound,''
  arXiv:0910.5347 [hep-th].
}

\lref\NojiriMH{
  S.~Nojiri and S.~D.~Odintsov,
  ``On the conformal anomaly from higher derivative gravity in AdS/CFT
  correspondence,''
  Int.\ J.\ Mod.\ Phys.\  A {\bf 15}, 413 (2000)
  [arXiv:hep-th/9903033].
}

\lref\Maldacena{
  J.~M.~Maldacena,
  ``The large N limit of superconformal field theories and supergravity,''
  Adv.\ Theor.\ Math.\ Phys.\  {\bf 2}, 231 (1998)
  [Int.\ J.\ Theor.\ Phys.\  {\bf 38}, 1113 (1999)]
  [arXiv:hep-th/9711200].
}

\lref\Witten{
  E.~Witten,
  ``Anti-de Sitter space and holography,''
  Adv.\ Theor.\ Math.\ Phys.\  {\bf 2}, 253 (1998)
  [arXiv:hep-th/9802150].
}

\lref\GKP{
  S.~S.~Gubser, I.~R.~Klebanov and A.~M.~Polyakov,
  ``Gauge theory correlators from non-critical string theory,''
  Phys.\ Lett.\  B {\bf 428}, 105 (1998)
  [arXiv:hep-th/9802109].
}

\lref\OsbornQU{
  H.~Osborn,
  ``N = 1 superconformal symmetry in four-dimensional quantum field theory,''
  Annals Phys.\  {\bf 272}, 243 (1999)
  [arXiv:hep-th/9808041].
}

\lref\HSa{
  M.~Henningson and K.~Skenderis,
  ``The holographic Weyl anomaly,''
  JHEP {\bf 9807}, 023 (1998)
  [arXiv:hep-th/9806087];
}
\lref\HSb{
  M.~Henningson and K.~Skenderis,
  ``Holography and the Weyl anomaly,''
  Fortsch.\ Phys.\  {\bf 48}, 125 (2000)
  [arXiv:hep-th/9812032].
}

\lref\BuchelSK{
  A.~Buchel, J.~Escobedo, R.~C.~Myers, M.~F.~Paulos, A.~Sinha and M.~Smolkin,
  ``Holographic GB gravity in arbitrary dimensions,''
  arXiv:0911.4257 [hep-th].
}

\lref\MetsaevTseytlin{
  R.~R.~Metsaev and A.~A.~Tseytlin,
  ``CURVATURE CUBED TERMS IN STRING THEORY EFFECTIVE ACTIONS,''
  Phys.\ Lett.\  B {\bf 185}, 52 (1987).
}

\lref\deBoerGX{
  J.~de Boer, M.~Kulaxizi and A.~Parnachev,
  ``Holographic Lovelock Gravities and Black Holes,''
  arXiv:0912.1877 [hep-th].
}

\lref\BlauVZ{
  M.~Blau, K.~S.~Narain and E.~Gava,
  ``On subleading contributions to the AdS/CFT trace anomaly,''
  JHEP {\bf 9909}, 018 (1999)
  [arXiv:hep-th/9904179].
}

\Title{\vbox{\baselineskip12pt
}}
{\vbox{\centerline{Supersymmetry Constraints in Holographic Gravities}
\vskip.06in
}}
\centerline{Manuela Kulaxizi${}^a$ and Andrei Parnachev${}^b$}
\bigskip
\centerline{{\it ${}^a$Department of Physics, University of Amsterdam }}
\centerline{{\it Valckenierstraat 65, 1018XE Amsterdam, The Netherlands }}
\centerline{{\it ${}^b$C.N.Yang Institute for Theoretical Physics, Stony Brook University}}
\centerline{{\it Stony Brook, NY 11794-3840, USA}}
\vskip.1in \vskip.1in \centerline{\bf Abstract}
\noindent
Supersymmetric higher derivative gravities define superconformal field theories via the AdS/CFT correspondence.
From the boundary theory viewpoint, supersymmetry implies a relation between the coefficients which
determine the three point function of the stress energy tensor which can be tested in the dual gravitational theory.
We use this relation to formulate a necessary condition for the supersymmetrization of higher derivative gravitational terms.
We then show that terms quadratic in the Riemann tensor do not
present obstruction to supersymmetrization, while generic higher order terms do.
For technical reasons, we restrict the discussion to seven dimensions where
the obstruction to supersymmetrization can be formulated in terms of the coefficients
of Weyl anomaly, which can be computed holographically.

\vfill

\Date{December 2009}


\noindent Higher derivative terms in gravity naturally arise
in the low energy limit of string theories.
It is interesting to investigate the role these terms
play in the AdS/CFT correspondence \refs{\Maldacena\Witten-\GKP}.
Common lore states that these terms encode corrections due to
deviation from the large $N$ or infinite coupling limit in the
boundary theory.
In fact, higher derivative terms are necessary if one wants
to study the holographic duals of four-dimensional CFTs  with different
$a$ and $c$ central charges of the conformal algebra.

Once these higher derivative terms are introduced, a natural
question is whether the gravity theory can be made supersymmetric.
An affirmative answer presumably implies that the dual boundary
theory is superconformal.
The two and three point functions of the stress
energy tensor of conformal field theories in $d>3$ dimensions
are completely specified by three coefficients $\AA,\BB,\CC$.
The superconformal Ward identity further reduces the number
of independent coefficients to two; the explicit form of the constraint
in four dimensions has been worked out in \OsbornQU.
In six dimensions the form of the constraint has been determined
in \deBoerPN, and it is not hard to generalize this  to arbitrary dimensions.
The three point functions of the stress energy tensor are related to
graviton scattering amplitudes in AdS.
(These on-shell amplitudes, and consequently, the results of this
paper, are unaffected by the field redefinitions
in the bulk.)
One should therefore be able to test whether there is an obstacle to making
a given gravity theory supersymmetric by computing these
scattering amplitudes and checking whether the constraint is satisfied\foot{It
would be interesting to generalize this to include other fields, but we leave this
for future work.}.

For technical reasons, we consider gravities in seven dimensions
which are dual to the six dimensional CFTs.
This is because the latter have a peculiar property; the three
coefficients in front of the B-type terms in the Weyl anomaly, which
we denote by $b_n, n=1,\ldots,3$,
are linearly related to $\AA,\BB,\CC$.
Hence, supersymmetry implies a linear relation between $b_n$.
More precisely, consider the Weyl anomaly in the form
\eqn\wanomaly{  \AA_W= E_6 + \sum_n^3 b_n I_n +\nabla_i J^i}
where $E_6$ is the Euler density in six dimensions, $I_n, n=1,\ldots3$
are three independent conformal invariants composed out of the Weyl tensor
and its derivatives, and the last term is a total derivative of a covariant expression.
The free field theory result for the B-type part of the Weyl anomaly is \BastianelliFTb
\eqn\anomalyfree{\eqalign{  b_1 &= {28\over3} n_s+{896\over3} n_f+{8008\over3}n_a \cr
                            b_2 &= {5\over3}n_s-32 n_f-{2378\over3} n_a    \cr
                            b_3  &= 2 n_s+40 n_f+180 n_a\cr}
}
where $n_s,n_a,n_f$ are the numbers of scalar fields, antisymmetric two-forms and
Dirac fermions in six dimensions.
In terms of free fields, the supersymmetry condition can be written as
\eqn\susyfree{ 6 n_a + n_s- 8 n_f=0  }
This defines a plane in the $(n_a,n_s,n_f)$ space which passes
through the origin, $(n_a=0,n_s=8,n_f=1)$ and the point $(n_a=1,n_s=10,n_f=2)$.
The former corresponds to a free scalar superfield in 6 dimensions while the latter
to two (2,0) multiplets.
Since the relations between $(n_a,n_s,n_f)$, $(\AA,\BB,\CC)$
and $(b_1,b_2,b_3)$ are linear, eq. \susyfree\ allows one to determine
the explicit form of the constraint that supersymmetry imposes on $b_n$.
The result is
\eqn\susyb{  b_1-2 b_2+6 b_3=0   }
In the following we are going to check if this constraint
is satisfied to leading order in the coefficients in front
of the higher derivative terms in the gravitational lagrangian.

Consider the following action with negative cosmological constant:
\eqn\action{S=\int \sqrt{-g} {\cal{L}}=\int \sqrt{-g}\left(R+{30 \over L^2}+\sum_i {\cal{L}}_i \right),\qquad
                   \LL_i=\sum_j a_{ij} \LL_{ij}}
In  \action\ ${\cal{L}}_i$ stand for all possible $\OO(R^i)$ higher derivative terms,
while $\LL_{ij}$ denotes all possible contractions of the Riemann tensors which are $\OO(R^i)$.
An AdS space  of length $L+\OO(a_{ij})$ is a solution of the equations of motion.
The explicit expressions for the higher derivative  terms ${\cal{L}}_2$ and ${\cal{L}}_3$ are
\eqn\ltwothree{\eqalign{{\cal{L}}_2 &=a_{21} R_{MNPQ}^2+ a_{22} R_{MN}^2+ a_{23} R^2\cr
{\cal{L}}_3 &= a_{31}  R^{IJKL}R_{KLMN} R^{MN}_{\quad IJ}+ a_{32}  R^{IJ}_{\quad KM} R^{KL}_{\quad JN} R^{MN}_{IL}+
a_{33} R^{IJKL}R_{KLJM}R^{M}_{\quad I}+\cr
&+ a_{34}  R R_{IJKL}^2+ a_{35}  R^{IKJL} R_{JI} R_{LK}+ a_{36}  R^{IJ}R_{JK}R^{K}_{\quad I} +a_{37} R R_{IJ}^2+ a_{38} R^3  }}
We will also consider an $\OO(R^4)$ term of the type
\eqn\lfour{  {\cal{L}}_4= a_{41}  R^{IJKL} R_{KLMN} R^{MNPQ} R_{PQIJ}+\ldots  }
In the following we will compute the leading corrections to the $b_n$
from all terms in \ltwothree, \lfour.
The leading (Einstein-Hilbert) result, $b_1^{(0)}=-1680, \quad b_2^{(0)}=-420, \quad b_3^{(0)}=140$  \refs{\HSa-\HSb}\
satisfies \susyb.
Each term in \ltwothree, \lfour\ is going to give rise to
\eqn\bexp{  b_n = b_n^{(0)} + a_{ij} \tb_n^{(ij)} + \OO( a_{ij}^2),\qquad n=1,\ldots,3}
Let us introduce
\eqn\Bdef{  B^{(ij)} = \tb_1^{(ij)}-2 \tb_2^{(ij)}+6 \tb_3^{(ij)}  }
The condition 
\eqn\susycondB{  \sum_{ij} a_{ij} B^{(ij)}\neq0  }
 implies that  there is an obstruction to the
supersymmetrization of the corresponding term in higher derivative gravity.

To compute $\tb^{(ij)}$ we will make use of the prescription developed in \refs{\HSa-\HSb}.
In practice, we will mostly follow \deBoerPN.
Consider the Einstein-Hilbert action with negative cosmological constant,
$a_{ij}=0$, and the following ansatz for the metric
\eqn\adsanz{  ds^2=L^2 \left({1\over 4\rho^2} d\rho^2+{1\over\rho} g_{\mu\nu}  dx^\mu dx^\nu \right) }
where
\eqn\gijexp{ g_{\mu\nu}=g_{\mu\nu}^{(0)}+\rho g_{\mu\nu}^{(1)}+\rho^2 g_{\mu\nu}^{(2)}+\rho^3 g_{\mu\nu}^{(3)}+\OO(\rho^3\log \rho)  }
is an expansion  in powers of the radial coordinate $\rho$.
One can now solve the Einstein equations of motion order by order in the $\rho$
expansion and  determine $g_{\mu\nu}^{(p)}, p=1,\ldots$ in terms
of $g_{\mu\nu}^{(0)}$.
The resulting expansion \gijexp\ is then substituted back into the Einstein-Hilbert action density  and
the coefficient of the $1/\rho$ term encodes the anomaly in Einstein-Hilbert  gravity.
To compute the $\OO(a_{ij})$ correction to the anomaly it is sufficient  to
evaluate the $\sqrt{\det g} \LL_{ij}$ term on the solutions
of the Einstein equations of motion and extract the $1/\rho$ term.
This is because the $\OO(a_{ij})$ contribution from the Einstein-Hilbert lagrangian
due to the $\OO(a_{ij})$ change in the solution \gijexp\ is proportional to the
equations of motion and vanishes on-shell\foot{A similar approach was used in \BlauVZ\ in the
context of $\OO(R^2)$ corrections to Einstein gravity in $AdS_5$.}.
In other words, we are going to compute
\eqn\weylan{ \sqrt{\det g^{(0)} } \AA_W^{(ij)}=
           \left[\sqrt{\det g} \LL_{ij}(g_{ij}) \right]_{1\over\rho} = \left[\sqrt{\det g^{(0)} }\AA_W^{(ij)} \right]_{g^{(0,1,2)}} +
\left[\sqrt{\det g^{(0)} }\AA_W^{(ij)} \right]_{g^{(3)}}   }
where $[\ldots]_{1\over\rho}$ means that we are extracting $1/\rho$ coefficient
from the expression in the square brackets.
In \weylan\ $\LL_{ij}$ is evaluated on the solution \gijexp\ of Einstein equations
of motion; for technical reasons it is convenient to separate the contributions
from $g^{(0,1,2)}$ and $g^{(3)}$.
In particular, the former piece
\eqn\waott{ \left[\sqrt{\det g^{(0)} }\AA_W^{(ij)} \right]_{g^{(0,1,2)}} =
        \left[\sqrt{\det g} \LL_{ij}\left(g=g^{(0)}+\rho g^{(1)}+\rho^2 g^{(2)}\right) \right]_{1\over\rho}    }
is evaluated on the metric truncated to the $\OO(\rho^2)$.
As in \deBoerPN\ we take the boundary metric to be of the form
\eqn\bdmetric{  g_{\mu\nu} dx^\mu dx^\nu = f(x^3,x^4) \left[ (dx^1)^2  +(dx^2)^2\right] +\sum_{i=3}^6 (dx^i)^2 }
and use Mathematica to determine $g^{(1)}$,
\eqn\gijone{    g^{(1)}_{\mu\nu}= -{1\over4}\left(R_{\mu\nu}-{1\over10} R g^{(0)}_{\mu\nu}\right)  }
and $g^{(2)}$ (which is slightly more complicated, so we do not quote it here).
In eq. \gijone\ $R_{\mu\nu}$ is the curvature tensor of the metric  $g^{(0)}$.
This way we completely determine \waott.

Unlike \deBoerPN, we also need an expression for $g^{(3)}$ which contributes
to the $\OO(a_{ij})$ term in the anomaly.
This is because we used the equations of motion to eliminate the correction
coming from the Einstein-Hilbert lagrangian.
It is easier to find $g^{(1)}$ and $g^{(2)}$ rather than $g^{(3)}$ because the $\OO(\rho^3\log \rho)$
in \gijexp\ contributes to the equations of motion at this order.
Fortunately, the contribution to $\tb_{ij}$ due to $g^{(3)}$
comes in a very simple form.
Namely, the  term in the anomaly \weylan\ due to $g^{(3)}$ is given by
\eqn\llgthree{   \left[\AA_W^{(ij)} \right]_{g^{(3)}} =  c_{ij}   \Tr\left[\left(g^{(0)}\right)^{-1} g^{(3)}\right] }
where $c_{ij}$ are easily found to be
\eqn\cija{  c_{21}= 9,\quad c_{22}=27,\quad c_{23}=189 }
\eqn\cijb{  c_{31}={-}6,\; c_{32}=6,\; c_{33}=18, \; c_{34}={-}126, \; c_{35}=c_{36}={-}54, \; c_{37}={-}378, \; c_{38}={-}2646  }
and $c_{41}=-12$.
To compute $\Tr\left[\left(g^{(0)}\right)^{-1} g^{(3)}\right]$
one can use equations of motion of Einstein-Hilbert gravity.
They have been written down in a convenient form in \HSa;
we only need the last line of equation (7) in \HSa:
\eqn\lastline{  \Tr\left[ g^{-1} g''\right] = {1\over2} \Tr\left[g^{-1} g' g^{-1} g'\right]  }
Here $g$ is the metric \gijexp\ and prime denotes differentiation with
respect to $\rho$.
Substituting the expansion \gijexp\ into \lastline\ one can use
the $\OO(\rho)$ term in the resulting expression to write
\eqn\gthree{\eqalign{  \Tr\left[ \left(g^{(0)}\right)^{-1} g^{(3)}\right]&=
     {1\over6}\Bigg(    4\Tr\left[  \left(g^{(0)}\right)^{-1}  g^{(1)} \left(g^{(0)}\right)^{-1}  g^{(1)} \right]\cr
              &\qquad -\Tr\left[  \left(g^{(0)}\right)^{-1} g^{(1)}\left(g^{(0)}\right)^{-1} g^{(1)}\left(g^{(0)}\right)^{-1}g^{(1)}\right]
              \Bigg)  \cr}
}
Together with \cija, \cijb\ and the solution for $g^{(1)}, g^{(2)}$, eq. \gthree\
allows us to compute \llgthree.
Combining this with \waott, we obtain an expression for the Weyl anomaly.
We then demand that the coefficient in front of every term  in the expression
\eqn\eqcoeff{ \AA_W^{(ij)}-\sum_{n=1}^3 \tb_n^{(ij)} I_n -\sum_{n=1}^7 c_n^{(ij)} C_n =0 }
vanishes.
In eq. \eqcoeff\ the $I_n$ are the B-type anomaly terms composed out of the Weyl
tensor, and $C_n$ are the total derivative terms.
Both can be found in Appendix A of \BastianelliFTb.
This completely fixes $\tb_n$ and $c_n$.
The results are summarized below (we omit
an overall coefficient common to all $\tb_n^{(ij)}$).
\eqn\bna{\eqalign{  \tb_1^{(21)}&={5\over96},\quad\tb_2^{(21)}= {37\over384},\quad \tb_3^{(21)}= {3\over128}; \qquad B^{(21)}=0   \cr
           \tb_1^{(22)}&={21\over32},\quad \tb_2^{(22)}= {21\over128},\quad \tb_3^{(22)}= -{7\over128}; \qquad B^{(22)}=0   \cr
           \tb_1^{(23)}&={147\over32},\quad\tb_2^{(23)}= {147\over128},\quad\tb_3^{(23)}= -{49\over128}; \qquad B^{(23)}=0   \cr}
}
Thus all $\OO(R^2)$ terms have vanishing $B$.
The results for the $\OO(R^3)$ terms are
\eqn\bnb{\eqalign{  \tb_1^{(31)}&={9\over16},\quad\tb_2^{(31)}= {9\over64},\quad \tb_3^{(31)}= -{41\over192}; \qquad B^{(31)}=-1   \cr
           \tb_1^{(32)}&={23\over16},\quad \tb_2^{(32)}= {7\over64},\quad \tb_3^{(32)}= -{31\over192}; \qquad B^{(32)}={1\over4}   \cr
           \tb_1^{(33)}&={5\over16},\quad\tb_2^{(33)}= {37\over64},\quad\tb_3^{(33)}= {9\over64}; \qquad B^{(33)}=0   \cr
           \tb_1^{(34)}&=-{35\over16},\quad\tb_2^{(34)}= -{259\over64},\quad \tb_3^{(34)}= -{63\over64}; \qquad B^{(34)}=0   \cr
           \tb_1^{(35)}&=-{63\over16},\quad \tb_2^{(35)}= -{63\over64},\quad \tb_3^{(35)}= -{21\over64}; \qquad B^{(35)}=0;
                                                                \qquad        \tb_n^{(35)}= \tb_n^{(36)}      \cr
           \tb_1^{(37)}&=-{441\over16},\quad \tb_2^{(37)}= -{441\over16},\quad \tb_3^{(37)}= {147\over64}; \qquad B^{(37)}=0;  \cr
           \tb_1^{(38)}&=-{3087\over16},\quad \tb_2^{(38)}= -{3087\over16},\quad \tb_3^{(38)}= {1029\over64}; \qquad B^{(38)}=0; \cr
}}
Finally,
\eqn\bnc{  \tb_1^{(41)}=-{25\over8},\quad\tb_2^{(41)}= -{89\over32},\quad \tb_3^{(41)}= {89\over96}; \qquad B^{(41)}=8   }
Let us discuss these results.
Apparently, there is no obstruction to supersymmetrizing
$\OO(R^2)$ terms, at least at the linear level.
This is consistent with the results of \deBoerPN\ where the supersymmetric
constraint for the Gauss-Bonnet term has been shown to hold.
It seems that this statement  might be dimension independent (the
analogous quantity has been shown to vanish in any dimensions \BuchelSK;
see also \HofmanMaldacena).

The situation with cubic terms is more interesting.
Apparently a generic term cubic in the Riemann tensor
cannot be supersymmetrized.
This is consistent with the fact that such terms
do not appear in superstring amplitudes \MetsaevTseytlin.
Note that it is possible to take a linear combination of the  $\OO(R^3)$ terms
to engineer $B=0$.
In fact, according to the recent results of \deBoerGX\
the cubic Lovelock term (Euler density in six dimensions)
is precisely of this type which implies
that there is no obstruction to supersymmetrizing this term.
Note that the Lovelock term vanishes upon dimensional reduction to four dimensions,
which is consistent with the expectations that $\OO(R^3)$ terms
cannot be supersymmetrized there\foot{We thank Martin Rocek
for explaining this point to us.}.
It is also interesting to observe that the generic $\OO(R^4)$ term
contributes to the anomaly, and hence to the three point function of the
stress energy tensor.
This contribution leads to  a non-vanishing value of $B$.

To summarize, we formulated the necessary condition for
supersymmetry in CFTs dual to higher derivative gravities.
Non-vanishing $B$ defined by \Bdef\ implies that the boundary
theory is not superconformal.
This can be used to check whether the relevant term in higher
derivative gravity can be supersymmetrized or not.
In particular, we found that $B=0$ for all terms of $\OO(R^2)$
but is generically nonvanishing for the $\OO(R^3)$ and  $\OO(R^4)$ terms.
We did not investigate $\OO(R^5)$ and higher derivative terms but see
no reason why they would generically lead to vanishing $B$.

\bigskip
\bigskip

\noindent {\bf Acknowledgement:}
We thank L. Rastelli, M. Rocek,  W. Siegel and M. Taylor for
useful discussions.
We also thank J. de Boer for collaboration on 
related projects.
The work of M.K. was partly supported by NWO Spinoza grant.

\footatend\vfill\supereject\immediate\closeout\rfile\writestoppt
\baselineskip=14pt\centerline{{\bf References}}\bigskip{\frenchspacing%
\parindent=20pt\escapechar=` \input refs.tmp\vfill\eject}\nonfrenchspacing
\end